\documentclass[letterpaper, 10 pt, conference]{ieeeconf}  

\IEEEoverridecommandlockouts                              

\usepackage{amsmath,amsfonts}
\usepackage{algorithmic}
\usepackage{algorithm}
\usepackage{array}
\usepackage[caption=false,font=normalsize,labelfont=sf,textfont=sf]{subfig}
\usepackage{textcomp}
\usepackage{stfloats}
\usepackage{url}
\usepackage{verbatim}
\usepackage{graphicx}
\usepackage{cite}

\usepackage{tikz}
\usepackage{graphics} 
\usepackage{epsfig} 
\usepackage{amssymb}  
\usepackage{url,algorithm}
\usepackage{multirow}
\usepackage{color}
\usepackage{adjustbox}

\newcommand{\rr}{{\mathbb R}}

\newcommand{\ba}[1]{\begin{array}{#1}}
	\newcommand{\ea}{\end{array}}

\usepackage{enumerate}

\title{\LARGE \bf Learning Critical Scenarios in Feedback Control Systems for Automated Driving}

\author{ Mengjia Zhu, Alberto Bemporad, Maximilian Kneissl, Hasan Esen 
	\thanks{M. Zhu and A. Bemporad are with IMT School for Advanced Studies Lucca,  Lucca, Italy. {\tt\small \{name.surname@imtlucca.it\}}}   %
	\thanks{M. Kneissl was with the Department of Corporate Research and Development, DENSO Automotive Deutschland GmbH, 85386 Eching, Germany, and is currently with Volvo Autonomous Solutions. {\tt\small maximilian.kneissl@volvo.com} } 
	\thanks{H. Esen is with the Department of Corporate Research and Development, DENSO Automotive Deutschland GmbH, 85386 Eching, Germany. {\tt\small h.esen@eu.denso.com} } 
}

\begin{document}

\maketitle
\thispagestyle{empty}
\pagestyle{empty}


\begin{abstract}
Testing is essential for verifying and validating control designs, especially in safety-critical applications. In particular, the control system governing an automated driving vehicle must be proven reliable enough for its acceptance on the market. Recently, much research has focused on scenario-based methods. However, the number of possible driving scenarios to test is in principle infinite. In this paper, we formalize a learning-based optimization framework to generate corner test-cases, where we take into account the operational design domain. We examine the approach on the case of a feedback control system for automated driving, for which we suggest the design of the objective function expressing the criticality of scenarios. Numerical tests on two logical scenarios of the case study demonstrate that the approach can identify critical scenarios within a limited number of closed-loop experiments.

\end{abstract}


\section{Introduction}
Advancements in algorithmic design, increasing experience through self-driving prototype vehicles, as well as first certifications for Level 4 automated vehicles~\cite{J3016} indicate rapid progress in the automated driving (AD) domain.
Safety argumentation of such highly automated vehicles plays a key role in introducing AD to open roads.
To achieve certification of AD systems, a proof that the vehicle operates safely must be provided.
To this end, verification and validation (V\&V) strategies are an essential stage for providing safety argumentations~\cite{SOTIF,saFAD}.

Due to the innumerable amount of possible scenarios that can happen during real-world driving, AD software architectures are designed to be continuously updatable even after system deployment.
To enable such updates, which might be safety critical, there is an intrinsic need for scalable V\&V methods. 
While coverage-driven V\&V strategies will play an important role for safety argumentation, it is of particular importance to be capable of extracting safety-critical test scenarios.

A large amount of research has focused on scenario-based methods~\cite{zhang2021finding,riedmaier2020survey}, for which an Operational Design Domain (ODD) is appropriately defined to limit the infinite scope of the applicability of the AD system. An ODD provides the set of conditions under which the AD system is designed to function~\cite{J3016}. By viewing a closed-loop test case as a scenario within an ODD, \emph{critical scenarios} are the test cases that can lead to harm.

Different exploration methods have been studied in the literature to find critical/corner test cases~\cite{zhang2021finding}. These methods can be broadly classified into \emph{na\"ive-search} (see, e.g.,~\cite{mullins2018adaptive,Gladisch2019-pk}) and \emph{guided-search} (see, e.g.,~\cite{akagi2019risk,zhou2017reduced,xia2017automatic}) methods. Because test cases are chosen independently during the na\"ive search, parallelizing the process can speed up the procedure. 
On the other hand, test-case simulations can be computationally expensive and time-consuming. Hence, it is preferred to use exploration methods that can reduce the number of simulation experiments. For instance, surrogate-based black-box optimization methods (guided-search methods) can be applied~\cite{Jon01,Bem20}. 
In addition, when the critical test cases are rare to identify (e.g., when the critical test cases are in a small region within the search domain), guided-search methods such as optimization and learning-based testing are often more experiment-efficient compared to na\"ive-search methods such as sampling and combinatorial testing methods~\cite{zhang2021finding}.

Nevertheless, most na\"ive search methods are applied when the other actors in the scenario do not involve parameter trajectories~\cite{zhang2021finding} because the search space can increase dramatically with versatile trajectories of other actors. In these cases, na\"ive-search methods become infeasible. Also, limited research has been devoted to using an optimization-based method to explore the scenario space that involves parametric trajectories~\cite{abeysirigoonawardena2019generating,zhang2021finding}. 

\subsection{Contribution}
In this paper, we formalize a learning-based optimization framework for determining safety-critical V\&V scenarios that aims at fast V\&V processes.
The optimization formulation ($i$) considers a definition of criticality, ($ii$) uses a formalized scenario space, ($iii$) finds critical scenarios quickly, and ($iv$) can explore the scenario space with and without parametric trajectories.

The main contributions are summarized as follows:
\begin{itemize}
\item A holistic formulation and solution strategy dealing with the posed V\&V problem, that takes an ODD description into account and translates the critical test scenario identification problem as an optimization problem with the optimization's objective function formulated as a criticality measure;

\item A detailed illustration of the overall approach when analyzing a feedback control system for AD, where we show how ODD can be selected to lower the number of optimization parameters when parametric trajectories are involved, and we design the objective function to identify collision occurrence;

\item Simulation results obtained by applying and evaluating the learning-based optimization strategy in two two-lane AD case studies, showing how the critical test scenarios identified can support the V\&V process also in early development phases.
\end{itemize}

 
The rest of this paper is organized as follows. Section~\ref{sec:problem} describes the problem formulation and the solution strategy. Following that, two case studies on automated driving are presented in Section~\ref{sec:case_study}. Lastly, conclusions and future research are discussed in Section~\ref{sec:conclusion}.

\section{Problem description and solution strategy}\label{sec:problem}
\begin{figure}[H]
  \centering
  \includegraphics[width=0.5\linewidth]{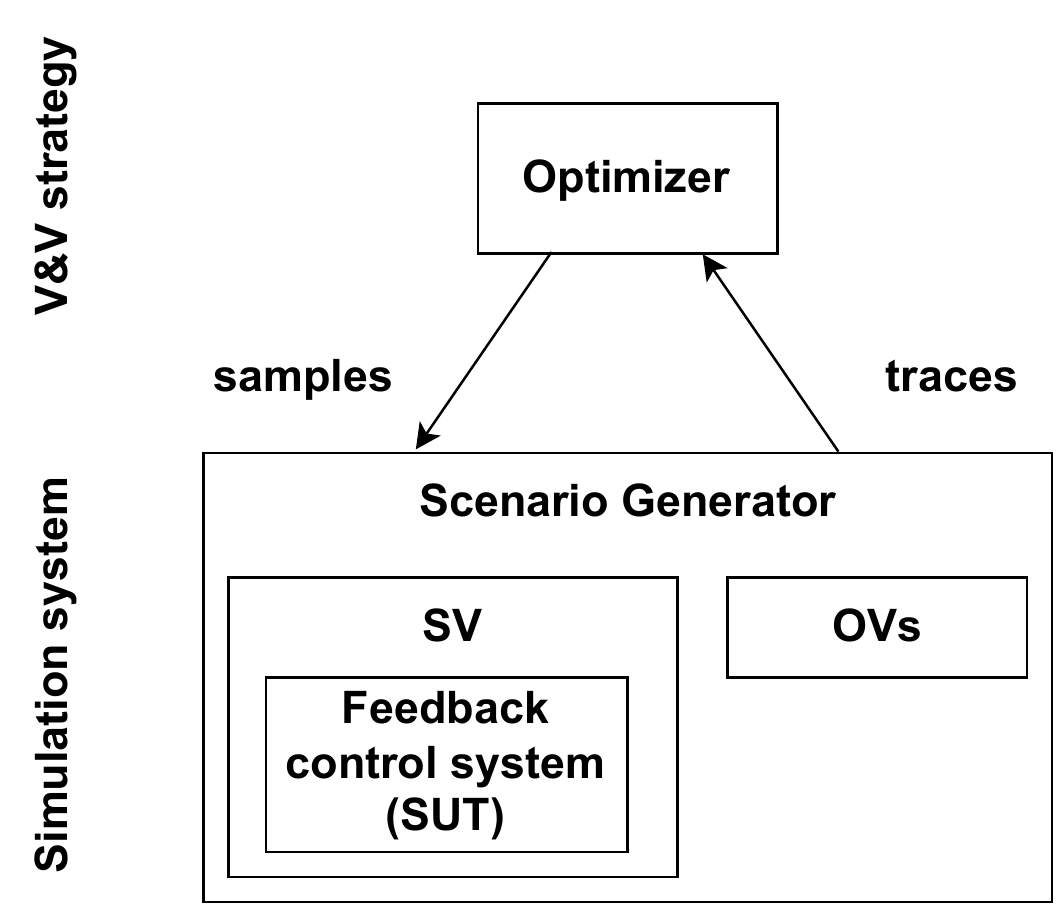}
  \caption{Simulation system and the V\&V strategy considered in this paper (SUT: System Under Test).}
  \label{fig:architecture_overall}
\end{figure}

To test the applicability of a designed feedback control system in an AD vehicle
we consider a simulation environment that consists of a subject vehicle (SV) and a given controller actuating it for lane keeping and to avoid collisions with different obstacle vehicles (OVs) (cf. Fig.~\ref{fig:architecture_overall}). We want to apply a scenario-based analysis in which each test scenario corresponds to
a particular behavior of the OVs.


To reduce test efforts, we propose to use a systematic way to efficiently identify testing scenarios for the V\&V of the designed feedback control system. In particular, we adopt and formalize the search-based testing framework and use learning-based optimization as the exploration method~\cite{Gladisch2019-pk,zhang2021finding} to find critical test scenarios for the control system, i.e., particular trajectories of the OVs that lead to collisions with the SV
when the latter is actuated by the given controller.

The overall architecture for the V\&V of the designed feedback control system (System Under Test, SUT) is summarized in Fig.~\ref{fig:architecture_overall}. In the following, we first introduce the general formulation of the optimization problem for critical test scenario identification. Then, we will describe a suitable optimization algorithm 
to solve the formulated problem. 

\subsection{General formulation of the optimization problem}
First, we need to define the ODD for the control system, which parametrizes a \emph{functional scenario} (a scenario space with semantic descriptions) into a \emph{logical scenario} (a state-space representation of the scenario space)~\cite{8500406,zhang2021finding}. Parameters of a logical scenario can then be further divided into \emph{assumed parameters} (APs) and \emph{parameters of interest} (PoIs), where PoIs can be constant or vary over time~\cite{zhang2021finding}. A particular scenario known as a \emph{concrete scenario} can be obtained by specifying values to each parameter (APs and PoIs) within the logical scenario~\cite{8500406,zhang2021finding}. We are interested in identifying the PoIs that lead to critical behaviors (\emph{e.g.}, short time-to-collision, and excessive jerk of the SV). In the following, we use $x_{\text{ODD}}\subseteq\mathbb{R}^n$ as the ODD parameters that define the logical scenario and ${x_{\text{scene}}\in x_{\text{ODD}}}$ as the scene parameters (PoIs) that describe snapshots of a logical scenario. For example, 
$x_{\text{scene}}$ may collect the initial distance between the subject and an obstacle vehicle and the acceleration of the latter.

For the control system under consideration, we translate the critical test scenario identification problem as the following optimization problem:
\begin{equation}\label{eq:opt_prob}
\begin{split}
x^*_{\text{scene}}\in \underset{x_{\text{scene}}}{\arg\min}\  &\quad f_{\text{system}}(x_{\text{scene}})\\
  \text{s.t.}  
                &\quad \ell \leq x_{\text{scene}} \leq u\\
               &\quad x_{\text{scene}} \in x_{\text{ODD}} \cap \chi,
\end{split}
\end{equation}
with the goal to identify a vector $x^*_{\text{scene}}$ of critical testing scenario parameters that can lead to critical behaviors. In~\eqref{eq:opt_prob}, $f_{\text{system}}: \rr^n \mapsto \rr$ can be a single assessing criterion (\emph{e.g.}, time-to-collision, distance between subject and obstacle vehicles), or a weighted combination of different
criteria; $\ell, u \in \rr^n$ are vectors of lower and upper bounds, and $\chi  \subseteq  \rr^n$ imposes further constraints on $x_{\text{scene}}$. Here, we assume that the objective function and the constraint formulations for~\eqref{eq:opt_prob} are known or pre-designed. We also stress that often a closed-form expression of $f_{\text{system}}$ with $x_{\text{scene}}$ is not available due to the complex way the level of criticality of the system depends on the variables in $x_{\text{scene}}$, although $f_{\text{system}}(x_{\text{scene}})$ can be evaluated through simulations.


\subsection{Optimization algorithm}
Surrogate-based optimization methods are suitable to solve~\eqref{eq:opt_prob}~\cite{RS13}. In this paper, the global optimization algorithm GLIS (GLobal optimization via Inverse distance weighting and Surrogate radial basis functions)~\cite{Bem20} is used to solve~\eqref{eq:opt_prob}. We summarize the main procedures of the GLIS algorithm in the following paragraphs and refer the reader to~\cite{Bem20} for a more detailed description. 

\begin{figure*}
  \centering
  \includegraphics[width=0.95\linewidth]{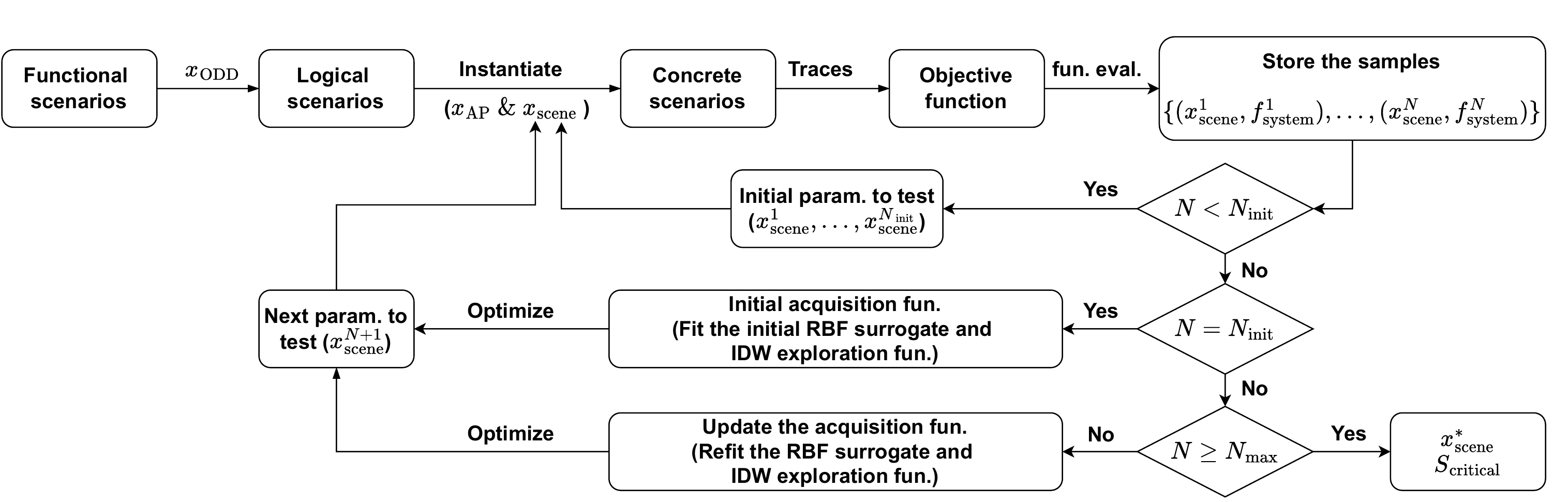}
  \caption{Optimization procedure of the GLIS algorithm ($S_{\text{critical}}$ refers to the set of vectors $x_{\text{scene}}$ that can lead to a critical test scenario).}
  \label{fig:glis_procedure}
\end{figure*}

The solution process of GLIS is divided into two phases: the initial sampling phase and the active learning phase. The optimization procedure of the GLIS algorithm for critical test scenario identification is summarized in Fig.~\ref{fig:glis_procedure}. A fixed computational budget $N_{\text{max}}$ is specified. In the initial sampling phase, GLIS runs $N_{\text{init}}$ experiments with different $x_{\text{scene}}$ sampled within the feasible domain. Traces from the generated concrete scenarios are fed to the optimizer and the objective function $f_{\text{system}}$ is evaluated (cf. Figures~\ref{fig:architecture_overall},~\ref{fig:glis_procedure}). At the end of the initial sampling phase (\emph{i.e.}, when $N = N_{\rm init}$), a surrogate function $\hat{f}_{\text{system}}$ is fitted to the initial samples $\{(x^1_{\text{scene}}, f^1_{\text{system}}), \dots, (x^{N_{\rm init}}_{\text{scene}}, f^{N_{\rm init}}_{\text{system}})\}$ using a radial basis interpolation function (RBF). In the active learning phase (\emph{i.e.}, when $N_{\text{init}}< N \leq N_{\text{max}}$), at each iteration, we query a new $x_{\text{scene}}$ parameter and update the surrogate function $\hat{f}_{\text{system}}$  by refitting $\hat{f}_{\text{system}}$ to the exiting and new samples $\{(x^1_{\text{scene}}, f^1_{\text{system}}), \dots, (x^N_{\text{scene}}, f^N_{\text{system}})\}$ using RBF interpolation.
Since solely minimizing $\hat{f}_{\text{system}}$ to find the next $x^{N+1}_{\text{scene}}$ to test may easily miss the global optimum, an inverse distance weighting (IDW) exploration function (weighted by an exploration parameter $\delta$) is summed with the RBF surrogate function to form an acquisition function ${a:\mathbb{R}^n\to\mathbb{R}}$. The acquisition function $a$ trades off exploitation of the RBF surrogate and exploration of the IDW function. We query the next point $x^{N+1}_{\text{scene}}$ for evaluation by minimizing function $a$ using a global optimization algorithm, \emph{e.g.}, Particle Swarm Optimization (PSO)~\cite{Ken10} or DIRECT~\cite{Jones2001},
which is a simple task since function $a$ is easy to evaluate. Then, a new concrete scenario is instantiated with $x^{N+1}_{\text{scene}}$, and the value $f_{\text{system}}(x^{N+1}_{\text{scene}})$ is evaluated. GLIS terminates when $N_{\text{max}}$ is reached. 

The main benefits of using GLIS are its easy incorporation of linear and/or nonlinear constraints and cheap computational cost~\cite{Bem20}. Additionally, the IDW exploration term in the acquisition function $a$ helps to efficiently explore the feasible domain. Moreover, one can tune the exploration parameter $\delta$ to obtain more focused or diverse results. On the other hand, one may select different optimization solvers depending on the particular formulation or prior knowledge of the objective function and/or constraints in~\eqref{eq:opt_prob}.
For example, Bayesian optimization (BO)~\cite{brochu2010} can be an alternative to GLIS. Comparisons between BO and GLIS on various numerical benchmarks are available in~\cite{Bem20}.

We remark an important aspect of the proposed method, related to the fact that we do not guarantee that the global minimum in~\eqref{eq:obj_fun_general} is achieved: if the best value $x^*_{\text{scene}}$
returned by the solver is not critical, this does not provide a guarantee that no critical
scenarios exist. This is a limitation of the proposed approach compared to formal method approaches~\cite{wing1990specifier} that aim instead at providing 100\% guarantees that all scenarios are safe. On the other hand, formal methods can be computationally prohibitive and/or over-conservative, while our approach can often quickly find critical scenarios, if they exist, which greatly helps the overall design and V\&V process.




\section{Automated driving case studies}\label{sec:case_study}
In this section, we apply the solution strategy proposed in Section~\ref{sec:problem} to two logical scenarios. Numerical tests are performed to identify relevant corner test cases for each logical scenario.
Here, we provide the system description of the SV and
a model predictive control (MPC) law governing it. A similar MPC design was considered in~\cite{zhu2020pref,zhu2021c} for preference-based calibration of the controller parameters, while for this study the controller design is fixed, \emph{i.e.}, we assume that the MPC parameters have been already calibrated. 

Computations are run on an Intel i7-8550U 1.8-GHz CPU laptop with 8GB of RAM. For GLIS, the Latin Hypercube Sampling (LHS) method~\cite{MBC79} (\emph{lhsdesign} function of  the  \emph{Statistics  and  Machine  Learning Toolbox} of MATLAB~\cite{matlabStaML}) is used in the initial sampling phase of GLIS~\cite{Bem20}, and Particle Swarm Optimization (PSO)~\cite{VV09} is used to minimize the acquisition function in the active learning phase. A multiquadric RBF is used to fit the surrogate model $\hat{f}_{\text{system}}$, with its hyperparameter $\epsilon$ set to 1 initially and recalibrated once at iteration $\lfloor N_{\text{init}}\text{+}(N_{\text{max}}\text{-}N_{\text{init}})/2\rfloor$. The exploration parameter $\delta$ in the acquisition function $a$ of GLIS is set to 2. As noted previously, one may use a larger $\delta$ to obtain more diverse results, at the price of possibly evaluating more combinations $x_{\text{scene}}$.

\subsection{Simulation model}\label{subsec:sys_des_sv}
\begin{figure}
  \centering
  \includegraphics[width=0.6\linewidth]{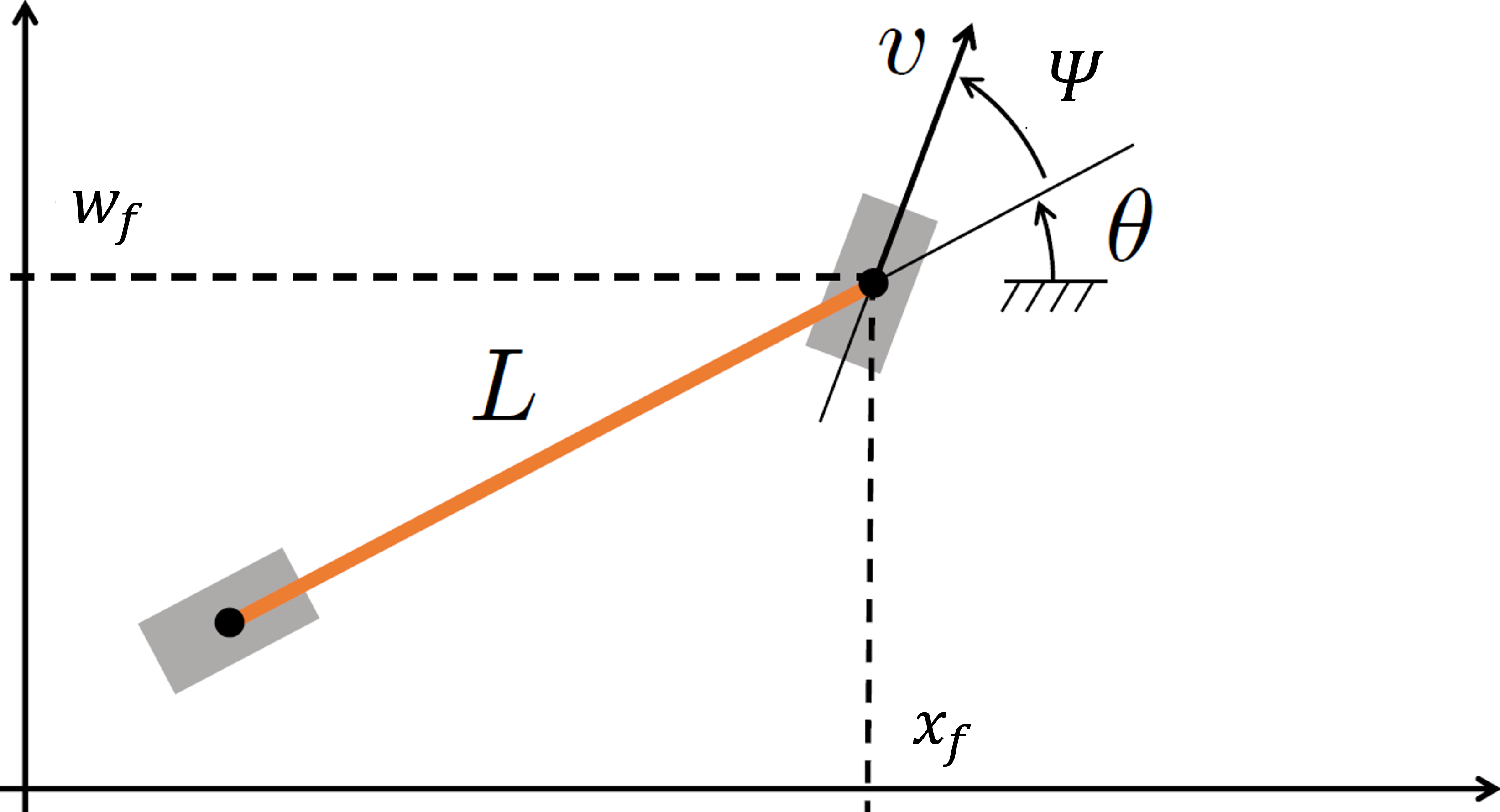}
  \caption{Two-degree-of-freedom bicycle model of the SV in Cartesian coordinates.}
  \label{fig:bicycle_model}
\end{figure}

We consider a simplified simulation environment with horizontal roads only and in which
the SV's kinematics is described by the following two-degree-of-freedom bicycle model in 
Cartesian coordinates (cf. Fig.~\ref{fig:bicycle_model})
\begin{equation}\label{eq:nonlin_bicyclemodel}
\begin{split}
 \dot{x}_f = & v \cos(\theta+\psi)\\
 \dot{w}_f = & v \sin(\theta+\psi)\\
 \dot{\theta} = & \frac{v\sin(\psi)}{L}
\end{split}
\end{equation}
where the longitudinal $x_f$ and lateral $w_f$ [m] positions of the front wheel determine the reference point of the SV, and $\theta$ [rad]
is the yaw angle, defining the three-dimensional state vector $s = [x_f \  w_f \  \theta]'$,
$L$ [m] is the SV length, and the velocity $v$ [m/s] and steering angle $\psi$ [rad] determine the vector of manipulated variables $u = [v \ \psi]'$.  

\subsection{MPC formulation}\label{subsec:mpc_form_sv}
We use the MPC described in~\cite{zhu2020pref,zhu2021c} for lane-keeping and obstacle-avoidance.

The SV is actuated by the designed MPC controller to keep the lane position and avoid collisions with obstacle vehicles (OVs). More specifically, when OVs are within a safety distance, the MPC controller commands the SV to change lane, decelerate, or accelerate, depending on the relative positions and other conditions. The ``Change lane'' decision is made when three conditions are met (noted in Algorithm~\ref{algo:const_mpc}). The first and third conditions are trivial and based on common driving practice. The second condition is used since for a lane change no additional constraints are imposed on the SV's velocity. When the second condition is not met, a collision is likely to be triggered during lane change. The decision making process detailed in Algorithm~\ref{algo:const_mpc} translates to output constraints imposed in the MPC formulation (cf. Equation (25) in~\cite{zhu2021c}).

\begin{algorithm}[ht!]
	\caption{Adaptive constraints of the MPC controller}
	\label{algo:const_mpc}
	~~\textbf{Input}: number of OVs ($k$); SV length ($L$); positions of SV and OVs at time step $t$: \{$(x_f^{\text{SV}}(t),w_f^{\text{SV}}(t)),(x_{f1}(t),w_{f1}(t)),\dots, (x_{fk}(t),w_{fk}(t))$\}; velocities of SV and OVs at $t$: \{$v^{\text{SV}}(t),v_1(t),\dots,v_k(t)$\}; safety distances: longitudinal ($x_{f,\text{safe}}$) and lateral ($w_{f,\text{safe}}$)
	\vspace*{.1cm}\hrule\vspace*{.1cm}
	\textbf{FOR} $i = 1,\dots, k$ \textbf{DO}
	\begin{enumerate}[~~~{}]
		\item \textbf{IF} SV and OV$_i$ are on the same lane and within safety distances (both longitudinal and lateral) \textbf{THEN}
		\begin{enumerate}[~~~{}]
		\item \textbf{IF} (OV$_i$ is ahead of SV) \&\& (no collision between SV and OV$_i$ will happen in the next step with the current velocity) \&\& (OV$_j$, $\forall j \neq i, i, j = 1,\dots,k $ are out of safety longitudinal ($x_{f,\text{safe}}$) and lateral distances ($w_{f,\text{safe}}$)) \textbf{THEN}:
			\begin{enumerate}[~~~{}]
				\item \textbf{Decision}: Change lane;
				\item \textbf{Update}:
				\begin{enumerate}[~~~{}]
					\item $\min$ $w_f^{\text{SV}} = w_{fi}$ + $w_{f,\text{safe}}$  \textbf{IF} change from lower lane to higher lane; \textbf{OR}  
					\item $\max$ $w_f^{\text{SV}} = w_{fi}$ - $w_{f,\text{safe}}$ \textbf{IF} change from higher lane to lower lane;
					\item (Note: `lower' and `higher' here refer to the relative lateral position of SV w.r.t OV$_i$)
				\end{enumerate}
				 
			\end{enumerate}
			\item \textbf{ELSE} 
			\begin{enumerate}[~~~{}]
				\item \textbf{Decision}: Decelerate or Accelerate;
				\item \textbf{Update}:
				\begin{enumerate}[~~~{}]
					\item $\min$ $x_f^{\text{SV}} = x_{fi}$ + $1.1L$ \textbf{IF} OV$_i$ is behind of SV; \textbf{OR} 
					\item $\max$ $x_f^{\text{SV}} = x_{fi}$ - $1.1L$ \textbf{IF} OV$_i$ is ahead of SV;
				\end{enumerate}
			\end{enumerate}
		\end{enumerate}
		\item \textbf{ELSE}
		\begin{enumerate}[~~~{}]
			\item \textbf{Decision}: Keep the default output constraints;
		\end{enumerate}		
		\end{enumerate}
		\textbf{END}.
	\vspace*{.1cm}\hrule\vspace*{.1cm}
	~~\textbf{Output}: MPC output constraints.
\end{algorithm}

In the following subsections, we discuss the application of the proposed solution strategy (Section~\ref{sec:problem}) to two different logical scenarios to identify relevant critical test cases for the MPC controller under test.

\subsection{Logical~Scenario~1}\label{sec:log_sce_1}
\begin{figure}
  \centering
  \includegraphics[width=0.9\linewidth]{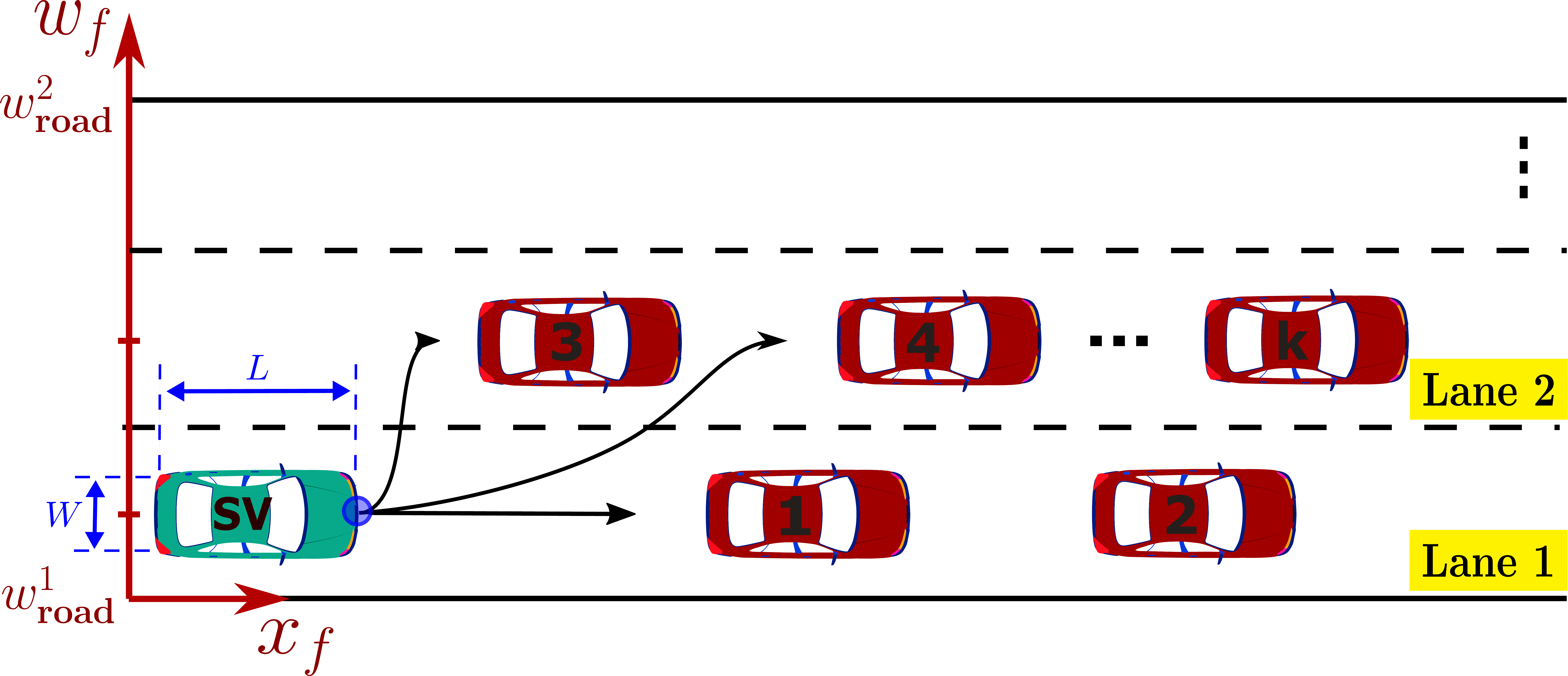}
  \caption{Logical~Scenario~1: the SV and OVs \{$1, \dots, k$\} are present on a one-way horizontal road with two or more lanes.}
  \label{fig:log_sce_1}
\end{figure}

A graphic representation of Logical~Scenario~1 is shown in Fig.~\ref{fig:log_sce_1}. In the following, we detail its ODD descriptions (APs and PoIs) that parametrize the functional scenario into the logical scenario.

\subsection*{\underline{ODD description}}
As shown in Fig.~\ref{fig:log_sce_1}, two or more vehicles are present on a one-way horizontal road with two or more lanes. The APs are the number of lanes, the road width, vehicle dimensions, and the experiment duration $t_{\text{exp}}$. OVs \{1,$\dots$,$k$\} can be placed on any lane, ahead or behind the SV. They move forward horizontally with a constant speed. We assume that no collision occurs among them. 
Here, the APs are the number of OVs ($k$), their initial lateral positions \{$w_{f1}^0,\dots,w_{fk}^0$\} and their constant yaw angles \{$\theta_1^0,\dots,\theta_k^0$\}. For Logical~Scenario~1, we define the PoIs as OVs' initial longitudinal positions \{$x_{f1}^0,\dots,x_{fk}^0$\} and initial velocities \{$v_1^0,\dots,v_k^0$\}. Different from OVs, the SV is commanded by a MPC controller to avoid collision as noted in Section~\ref{subsec:mpc_form_sv}. The APs for the SV are its initial longitudinal and lateral positions ($x_f^{\text{SV},0},w_f^{\text{SV},0}$) and parameters and constraint values in the MPC controller (cf. Section~\ref{subsec:mpc_form_sv}).

\subsection*{\underline{Numerical tests}}
We run three numerical tests on Logical~Scenario~1 to identify corner test cases. For all three tests, we set the road width to 6 m with 2 lanes (3 m/lane) and vehicle dimensions as $L = 4.5$ m and $W = 1.8$ m for both the SV and OVs. The safety longitudinal ($x_{f,\text{safe}}$) and lateral ($w_{f,\text{safe}}$) distances are specified to  10 m and 3 m, respectively. The lateral coordinates of the road ($w^1_\text{road}$ \& $w^2_\text{road}$ in Fig.~\ref{fig:log_sce_1}) are set to $\text{-}1.5$ and 4.5 m with SV initially placed at the center of Lane~1, \emph{i.e.}, $w_f^{\text{SV},0} = 0$ and $w^{\text{SV}}_f \in [\text{-}0.6,3.6]$. The longitudinal positions of both SV and OVs are considered as relative to $x_f^{\text{SV},0}$ with $x_f^{\text{SV},0} = 0$. For OVs, their constant yaw angles are set to 0$^\circ$, while $k$ and \{$w_{f1}^0,\dots,w_{fk}^0$\} are specified in each numerical tests (Table~\ref{tab:num_test_spec_logSec_1}). The optimization variables (PoIs) are $x_{\text{scene}}= [x^0_{f1},v^0_{1},\dots,x^0_{fk},v^0_{fk}]$.

\begin{figure}
  \centering
  \includegraphics[width=0.7\linewidth]{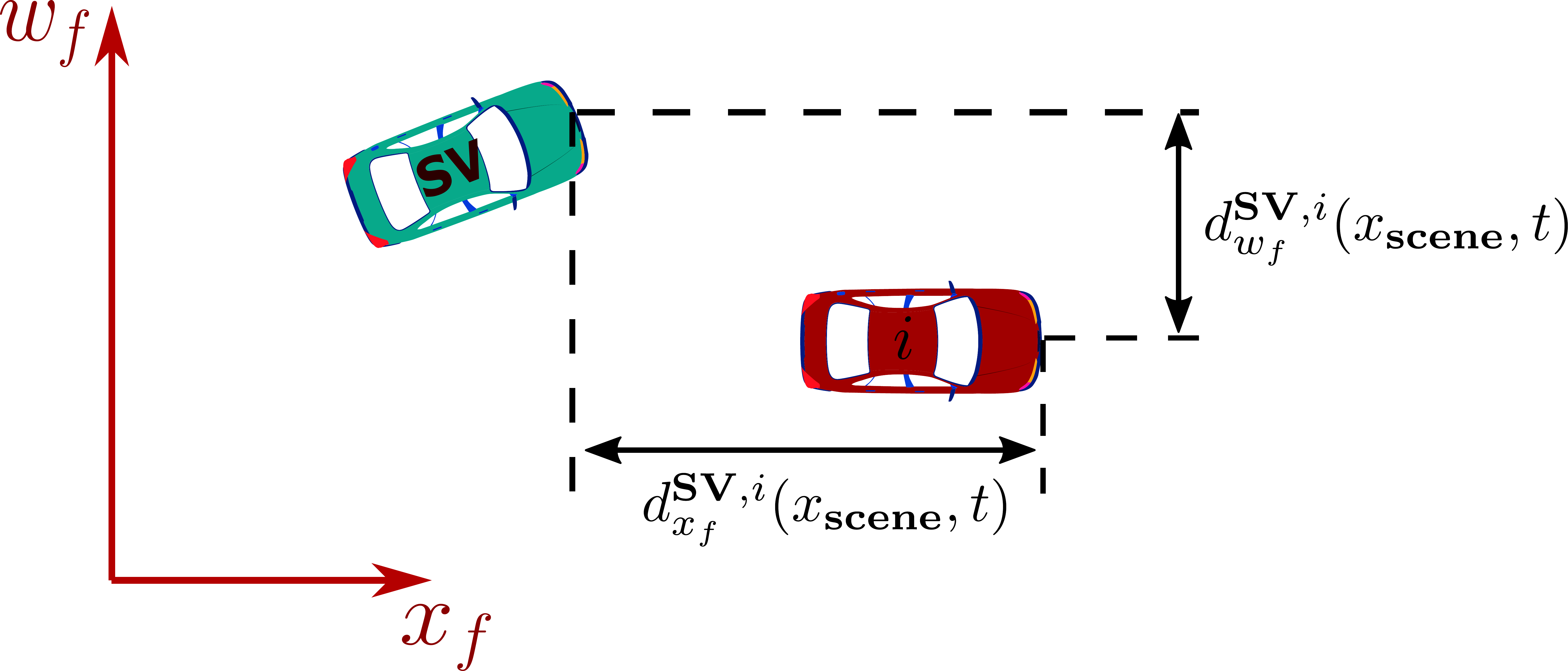}
  \caption{Longitudinal and lateral distances between SV and OV$_i$.}
  \label{fig:obj_defn_k}
\end{figure}

The objective function $f_{\text{system}}$ in~\eqref{eq:opt_prob} is required to drive the optimization process. Depending on the critical scenarios of interest, different surrogate measures may be used to assess criticality~\cite{zhang2021finding}. Furthermore, single or multiple criteria can be introduced in the objective function. In this paper, we propose the following objective function based on collision occurrence by analyzing the relative positions between the SV and OV$_i$ at different time step $t$:

\begin{equation}\label{eq:obj_fun_general}
\begin{split}
   \underset{x_{\text{scene}} \in x_{\text{ODD}}}{\min} & \quad \sum_{i = 1,\dots,k} d^{\text{SV},i}_{{x_f},{\text{critical}}}(x_{\text{scene}}) +d^{\text{SV},i}_{{w_f},{\text{critical}}}(x_{\text{scene}})\\ 
   \quad \text{s.t.} & \quad \ell \leq x_{\text{scene}} \leq u \ \ \&\ \text{other constraints} \\
    \text{where} & \quad d^{\text{SV},i}_{{x_f},{\text{critical}}}(x_{\text{scene}}) =\\
  & \quad \quad 
  	\begin{cases} 
  	\underset{t \in T_{\text{collision}}}{\min} d^{\text{SV},i}_{x_f}(x_{\text{scene}},t) &  \mathcal{I}_{\text{collision}}^i \\
  	 L & \sim\mathcal{I}_{\text{collision}}^i \&\ \mathcal{I}_{\text{collision}}\\
   \underset{t \in T_{\text{sim}}}{\sum} d^{\text{SV},i}_{x_f}(x_{\text{scene}},t) &  \sim \mathcal{I}_{\text{collision}}   
   \end{cases}  \\ 
  & \quad d^{\text{SV},i}_{{w_f},{\text{critical}}}(x_{\text{scene}})= \\
  & \quad \quad 
  \begin{cases}  
  \underset{t \in T_{\text{collision}}}{\min} d^{\text{SV},i}_{w_f}(x_{\text{scene}},t) &  \mathcal{I}_{\text{collision}}^i\\
    w_{f,\text{safe}} & \sim\mathcal{I}_{\text{collision}}^i \&\ \mathcal{I}_{\text{collision}}\\
    \underset{t \in T_{\text{sim}}}{\sum} d^{\text{SV},i}_{w_f}(x_{\text{scene}},t) &  \sim \mathcal{I}_{\text{collision}} 
  \end{cases}  \\
  &\quad \mathcal{I}_{\text{collision}}^i= \  \text{True},\ \text{if}\ \exists t\in T_{\text{sim}},\  \text{s.t.}\\
 & \quad (d^{\text{SV},i}_{x_f}(x_{\text{scene}},t) \leq L)\  \&\ (d^{\text{SV},i}_{w_f}(x_{\text{scene}},t) \leq W),\\
   & \quad \mathcal{I}_{\text{collision}} = \  \text{True},\ \text{if}\ \exists h \in \{1, \dots, k\}, \  \text{s.t.}\\
  & \quad \mathcal{I}_{\text{collision}}^h=\  \text{True}.
\end{split}
\end{equation}
Here, $d^{\text{SV},i}_{{x_f},{\text{critical}}}(x_{\text{scene}})$ and $d^{\text{SV},i}_{{w_f},{\text{critical}}}(x_{\text{scene}})$ are the critical longitudinal and lateral distances between SV and OV$_i$, respectively, while $d^{\text{SV},i}_{x_f}(x_{\text{scene}},t)$ and $d^{\text{SV},i}_{w_f}(x_{\text{scene}},t)$ are the longitudinal and lateral distances between SV and OV$_i$ at time step $t$ as shown in Fig.~\ref{fig:obj_defn_k}. $T_{\text{sim}}$ is the set of all time steps recorded during the experiment and
$T_{\text{collision}} \subseteq  T_{\text{sim}}$ is the set of time steps during the experiment where a collision occurs. $\mathcal{I}_{\text{collision}}$ is the indicator function for collision occurrence: $\mathcal{I}_{\text{collision}} = $ True if collision between SV and any OV occurs at any time step $t \in T_{\text{sim}}$; $\mathcal{I}_{\text{collision}} = $ False otherwise. $\mathcal{I}^i_{\text{collision}}$ is the indicator function for collision occurrence between SV and OV$_i$. 

We note some rationales for the formulation of the objective function~\eqref{eq:obj_fun_general}. When no collision occurs between SV and OV$_i$, \emph{i.e.}, when $\mathcal{I}^i_{\text{collision}} = $ False, constant values are assigned to the critical longitudinal and lateral distances. It is because the magnitudes of the corresponding longitudinal or lateral distances are irrelevant concerning criticality (collision occurrence in this case) when the controller can avoid collision under the defined ODD conditions. Therefore, including the actual distance measurements for non-collision cases will not help the solution process and are more likely to mislead the optimizer to find cases where SV and OV$_i$ drive side-by-side at close distances. 
However, if no collision occurs between SV and any OV in the scenario (including OV$_i$),  \emph{i.e.}, when $\mathcal{I}_{\text{collision}} = $ False, assigning constant values to every OV makes all the no-collision cases indistinguishable and provides limited information to the optimizer to guide the search. Therefore, when $\mathcal{I}_{\text{collision}} = $ False, the critical longitudinal and lateral distances of each OV$_i$ are set to the sum of its longitudinal and lateral distances at every time step, respectively. Minimizing the distances between SV and each OV$_i$ throughout the experiments increases the chance of collision occurrence. Since the magnitude of the sum is significantly higher than the constant distance assigned (when $\mathcal{I}^i_{\text{collision}} = $ False) and the minimum distance identified during a collision between SV and any OV (when $\mathcal{I}_{\text{collision}} = $ True), it can help guide the search during the optimization process without outweighing the collision cases. We also stress that depending on the criticality interested, one can blend the critical distances differently or use an alternative function $f_{\text{system}}$ to guide the search in the optimization process.

The problem specifications for the numerical tests are presented in Table~\ref{tab:num_test_spec_logSec_1}. The number of initial samples ($N_{\text{init}}$) used in GLIS is chosen to be $\lceil N_{\text{max}}/4 \rceil$. Every closed-loop experiment in each numerical test is simulated for 30 s ($t_{\text{exp}}$). Other constraints are used to prevent collisions among OVs. 

To show the efficiency of the proposed approach, we compare it with a random sampling method. The random samples are generated using the LHS method as in the initial sampling phase of GLIS. We run a Monte Carlo simulation with 20 runs of GLIS and LHS sampling methods to obtain statistically significant results.

\begin{table*}[!bt]
	\caption{Numerical Test - Problem Specification (Logical~Scenario~1)}\label{tab:num_test_spec_logSec_1}
	\centering
	\begin{adjustbox}{width=1\textwidth}
	\begin{tabular}{l|lllllll}
		\hline
		Test & $k$ & \{$w_{fi}^0$\} & $x_{\text{scene}}$ & $\ell$ and $u$ & Other constraints &  $N_{\text{max}} $& $N_{\text{init}} $ \\
		\hline
		
		1 &	1 & \{0\}			& $[x^0_{f1},v^0_{1}]'$	 & $\ell = $ [5,30]'; & N/A  & 50 & 13\\
		\vspace{0.2em}
		&	 & 			& 								 & $u = $ [50,80]'\\ 
		
		2 & 3 &	\{0,3,3\}		& $[x^0_{f1},v^0_{1},\dots,x^0_{f3},v^0_{f3}]'$	 &$\ell = $ [15,30,0,10,10,30]'; & $x^0_{f3} - x^0_{f2} > L$; & 100 & 25\\
		\vspace{0.2em}
		&	 & 			& 								 & $u = $ [50,80,100,80,100,80]' & $v^0_{3} > v^0_{2} $\\
		3 &	5 &	\{0,0,3,3,3\}	& $[x^0_{f1},v^0_{1},\dots,x^0_{f5},v^0_{f5}]'$ &$\ell = $ [15,30,0,10,0,10,10,10,20,10]'; & $x^0_{f2} - x^0_{f1} > L$; $x^0_{f4} - x^0_{f3} > L$; $x^0_{f5} - x^0_{f4} > L$; &  100 & 25\\
		\vspace{0.2em}
		&	 & 			& 								 & $u = $ [50,80,100,80,100,80,100,80,100,80]' & $v^0_{2} > v^0_{1} $; $v^0_{4} > v^0_{3} $; $v^0_{5} > v^0_{4} $ \\
		\hline
	\end{tabular}
	\end{adjustbox}
\end{table*}

\subsection*{\underline{Results}}
The average numbers of collision cases identified by GLIS and random sampling methods for different tests and their 95\% confidence interval are displayed in Table~\ref{tab:compare_random}. As shown in the table, the proposed framework can identify collision cases more frequently than LHS sampling.

\begin{table}[!bt]
\caption{Efficiency Comparison (Logical~Scenario~1)}\label{tab:compare_random}
\begin{tabular}{llll}
\hline
\multicolumn{1}{l|}{\multirow{2}{*}{Sampling method}} & \multicolumn{3}{l}{Collision Occurence} \\ \cline{2-4}
\multicolumn{1}{l|}{}             & Test 1 & Test 2 & Test 3 \\ \hline
\multicolumn{1}{l|}{GLIS}         &  $ 4 \pm 0 $      &  $ 30 \pm 12 $      &       $ 32 \pm 14 $ \\
\multicolumn{1}{l|}{Random sampling (LHS)} &   $0 \pm 0 $     &  $ 2 \pm 1 $      &       $4 \pm 1$ \\ \hline
\multicolumn{4}{l}{\text{The numbers are rounded to the nearest integers. }}                               
\end{tabular}

\end{table}

In the following, we present the results for one run of the Monte Carlo simulation for each test to analyze the collision cases. The optimization results for the numerical tests are shown in Table~\ref{tab:num_test_results_logSec_1}. In general, within a small number of experiments, GLIS can find critical test scenarios under the defined ODD conditions of the MPC controller of SV under testing. These identified critical scenarios often reveal features of a scenario that can lead to a collision. This information can then be used to support controller design and ODD refinement. 

\begin{table}[!bt]
\caption{Numerical Test - Results (Logical~Scenario~1)}\label{tab:num_test_results_logSec_1}
\begin{adjustbox}{width=0.48\textwidth}
\begin{tabular}{l|ll}
\hline
Test               & Iter & $x_{\text{scene}} = [x^0_{f1},v^0_{1},\dots,x^0_{fk},v^0_{fk}]$                                   \\ \hline

\multirow{3}{*}{1} & 18   &       [5 41.72]'                                \\

                   & 19   &       [5 36.62]'                                  \\ 
                   \vspace{0.3em}
                   & 21$^*$   &     [5 30.89]'                                    \\ 
   
\multirow{3}{*}{2} & 51$^*$   &   [15.00	30.00	44.14	10.00	49.10	47.39]'            \\
                   & 79    &   [28.09	30.00	70.29	10.00	74.79	31.74]'      \\
                    \vspace{0.3em}
                   & 40   &    [34.30	30.00	60.59	10.00	77.80	35.97]'             \\ 
                   
\multirow{3}{*}{3}	& 75$^*$   & [15.00	30.00	19.50	30.01	48.54	10.00	60.32	10.00	86.32	51.26]'   \\ 
					& 97   & [22.89	30.00	57.34	30.00	56.06	10.00	68.76	24.45	73.26	41.54]' \\
					& 76    &  [29.46	30.00	62.40	36.42	42.87	16.84	65.56	31.00	76.14	42.29]' \\ \hline
\multicolumn{3}{l}{Three sample iterations that can lead to collision are shown on the table.}\\
\multicolumn{3}{l}{(The ones with  $^*$ are the `best'/most critical ones identified by the optimizer }\\
\multicolumn{3}{l}{among these collision scenarios.)}\\
\end{tabular}
\end{adjustbox}
\end{table}

\begin{figure}
  \centering
  \includegraphics[width=0.9\linewidth]{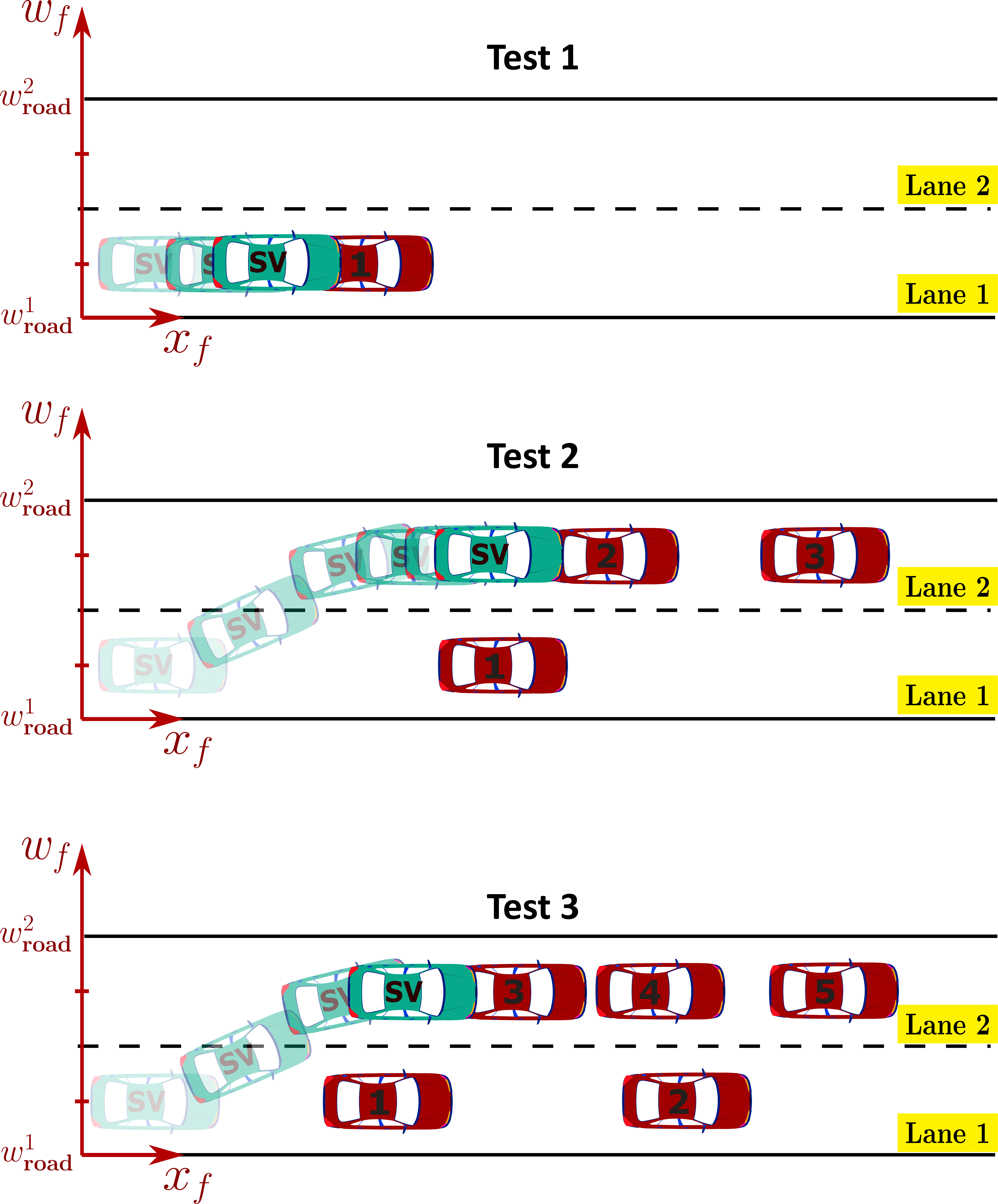}
  \caption{Collision cases for Logical~Scenario~1.}
  \label{fig:collision_log_sec_1}
\end{figure}

For Test 1, GLIS identifies 4 collision cases within 50 simulation experiments. Three sample iterations that can lead to collision are shown on Table~\ref{tab:num_test_results_logSec_1}. The collision triggering conditions in this test are small $x^0_{f1}$ and $v^0_1$, more specifically, when both are close to their lower bounds. With this group of $x_{\text{scene}}$, SV is not able to brake fast enough to avoid collision with OV$_1$ (Fig.~\ref{fig:collision_log_sec_1}). To eliminate these collision cases, one needs to either update the controller design (\emph{e.g.}, incorporate larger deceleration rate or use dynamic safety distance metrics) or refine the ODD definition of the applicability of the controller (\emph{e.g.}, update the lower bound of $x^0_{f1}$ or $v^0_1$ or both). 

For Test 2, GLIS finds 64 collision cases within 100 simulation experiments. In all critical cases, a collision occurs between SV and OV$_2$. Common features of the cases that triggered a collision are summarized here: ($i$) a relatively large $x_{f1}^0$ coupled with a relatively slow $v^0_1$ and the smaller $x_{f1}^0$, the greater $v^0_1$ (Table~\ref{tab:num_test_results_logSec_1}); ($ii$) a slow $v_2^0$ with a large $x_{f2}^0$. We note that the exact values of $x^0_{f1}$, $v^0_1$, $x_{f2}^0$, and $v_2^0$ leading to a collision depend on each other since the simulated experiment is dynamic. Nevertheless, the combinations that lead to a collision have the simulation conditions that cause the MPC controller to have the following decision-making process: the MPC controller first commands SV to change the lane to avoid collision with OV$_1$ on Lane~1, and then to decelerate to avoid OV$_2$ on Lane~2 (Fig.~\ref{fig:collision_log_sec_1}). However, SV cannot decelerate fast enough, leading to a collision with OV$_2$. Note that lane change is not an option in this case since OV$_1$ on Lane~1 blocks the way. To fix these critical scenarios, we need to update either the controller design or the ODD definitions, as noted in Test 1.

\begin{figure}
  \centering
  \includegraphics[width=0.85\linewidth]{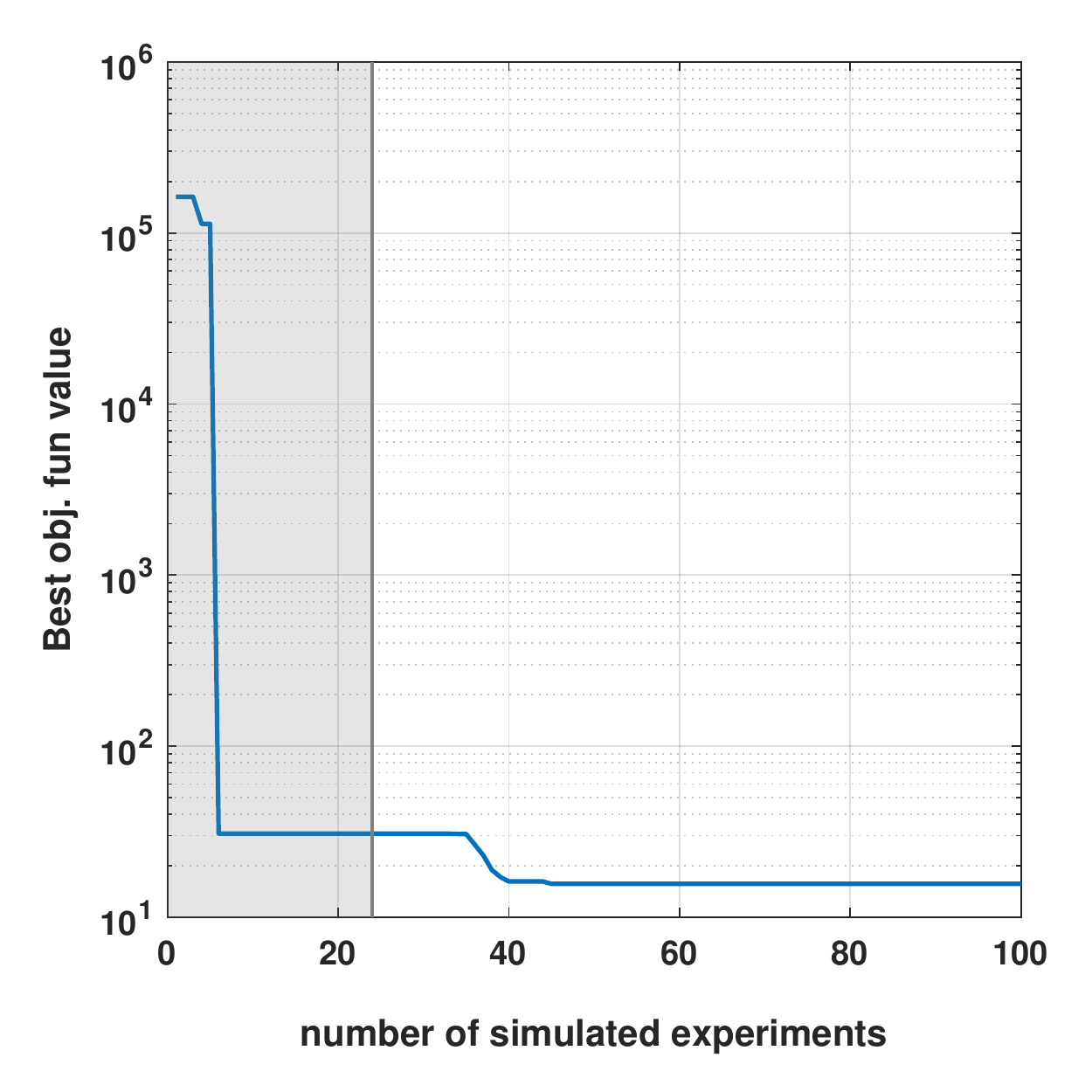}
  \caption{Best objective function value obtained as a function of the number of simulated experiments for Test 3 of Logical~Scenario~1. The vertical line denotes the last experiment in the initial sampling phase of GLIS.}
  \label{fig:log1_test3_eval}
\end{figure}

For Test 3, 73  collision cases are identified by GLIS in 100 simulation experiments. Fig.~\ref{fig:log1_test3_eval} shows the best objective function value obtained as a function of the number of simulated experiments. The collision happens between SV and OV$_3$. The collision triggering conditions of these scenarios are similar to the ones identified in Test 2. Initially, SV changes the lane to avoid OV$_1$. After switching lane, SV collides with OV$_3$ (Fig.~\ref{fig:collision_log_sec_1}). The collision is inevitable because OV$_3$ ahead of SV moves slowly, and SV cannot brake fast enough. Furthermore, depending on the initial conditions of OVs, either OV$_1$ or OV$_2$ or both block the way for a lane change of SV. A similar remedy plan as the previous two tests can eliminate these collision cases.

Overall, the three numerical tests demonstrate the capability of the proposed method to identify critical test scenarios of the designed controller under the defined ODD conditions. For Logical~Scenario~1, where the movement of OVs are restricted, by comparing Test 2 and 3, we observe that adding more OVs in the scenario does not provide more insights for potential critical scenarios. It is because SV only interacts with surrounding OVs and its obstacle avoidance mechanism with each OV is the same. However, Test 3 demonstrates the ability of GLIS to handle relatively high dimensional problems. 

As a side observation, we noticed that when running the numerical tests on the initial implementation of the MPC controller, GLIS occasionally drove the search to ``fake`` critical test scenarios attributed to some mis-implementation (bugs in the code) of the MPC controller. Such initial critical scenarios were very useful to quickly detect and fix these implementation errors that were nontrivial to identify a priori. Therefore, the cyclic interaction between GLIS obtaining results and the designer interpreting them and altering the design accordingly can significantly facilitate the V\&V cycle to identify actual corner cases within the defined ODD conditions. 

\subsection{Logical~Scenario~2}\label{sec:log_sce_2}
\begin{figure}
  \centering
  \includegraphics[width=0.9\linewidth]{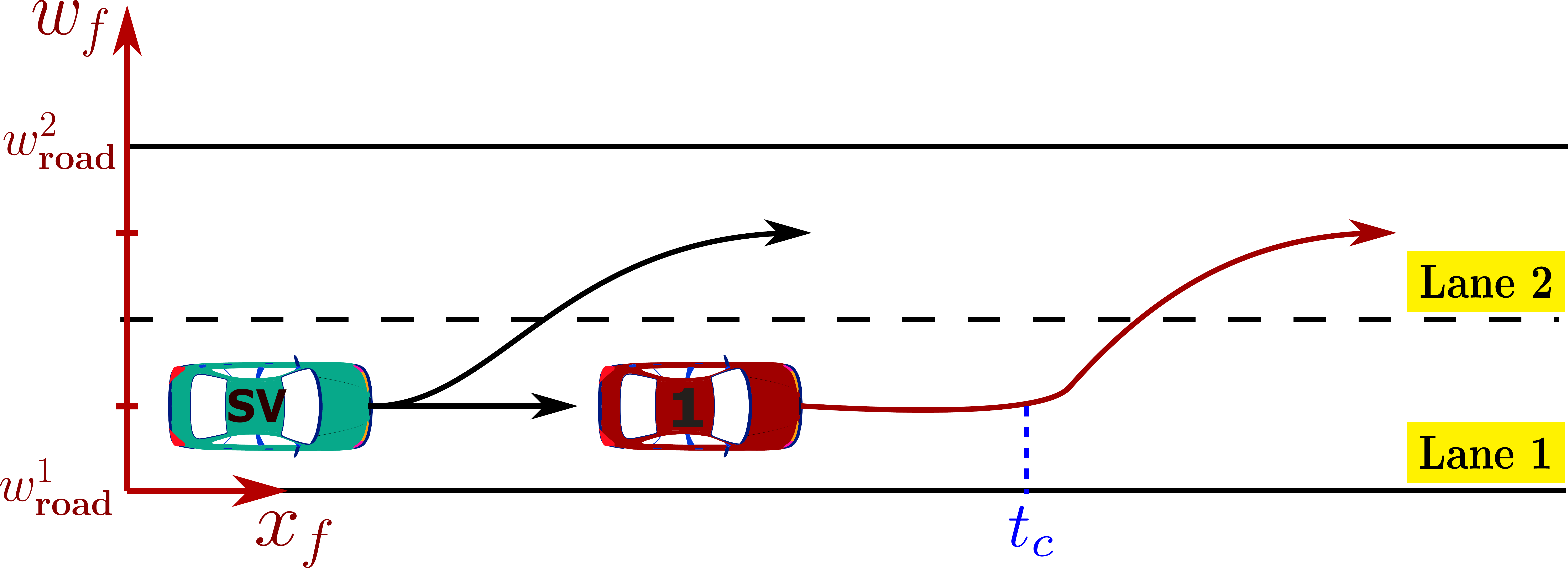}
  \caption{Logical~Scenario~2: the SV and one OV are present on a one-way horizontal road with two lanes.}
  \label{fig:log_sce_2}
\end{figure}

In the following, we discuss Logical~Scenario~2 (Fig.~\ref{fig:log_sce_2}) and the numerical test performed. We note that in Logical~Scenario~2, the trajectory of the OV varies and parametric trajectories are involved. 

\subsection*{\underline{ODD description}}
The ODD definitions of Logical~Scenario~2 are similar to Logical~Scenario~1, with the difference noted as the following: ($i$) the road only has two lanes; ($ii$) only one OV on the road; ($iii$) OV$_1$ is also commanded by an MPC controller to change the lane. Here, the MPC controller is used to obtain realistic trajectories of OV$_1$ and, at the same time, keep the dimension of $x_{\text{scene}}$ low.
For Logical~Scenario~2, the APs for SV remain the same as in Logical~Scenario~1. For OV$_1$, it is initially placed ahead of SV on Lane~1 and moves forward horizontally with a constant speed until a switching time, after which an MPC controller commands OV$_1$ to change from Lane~1 to Lane~2 at a constant speed (Fig.~\ref{fig:log_sce_2}). In this case, the APs for OV$_1$ are its initial lateral position ($w_{f1}^0$), its constant yaw angle ($\theta_1^0$) before switching, and the reference velocity and reference yaw angle for MPC controller after switching. We select the PoIs as its initial longitudinal position ($x^0_{f1}$), initial velocity ($v_1^0$) and switching time ($t_c$). The ODD of Logical~Scenario~2 is defined as such to provide a variety of OV trajectories while also reducing the number of required optimization parameters.

\subsection*{\underline{Numerical tests}}
We run one numerical test on Logical~Scenario~2. The controller under test is the MPC controller for SV as detailed in Section~\ref{subsec:mpc_form_sv}. The road and vehicle dimensions and the SV's initial conditions are set to the same values as in Logical~Scenario~1. 
Another MPC controller is used to command OV$_1$ for the lane-changing task after $t_c$. We emphasize that the MPC controller for OV$_1$ is only designed for lane-changing without a collision avoidance capability.

The objective function~\eqref{eq:obj_fun_general} is used with $k =1$. Here, the optimization variables (PoIs) $x_\text{scene} = [x^0_1, v^0_1, t_c]$, with $\ell = [11,30,0]'$ and $u = [50,80,40]'$. For GLIS, $N_{\text{max}}$ is set to 100 with $N_{\text{init}} = 25$, and each experiment is simulated for 30 s ($t_{\text{exp}}$).

\subsection*{\underline{Results}}
\begin{figure}
  \centering
  \includegraphics[width=0.9\linewidth]{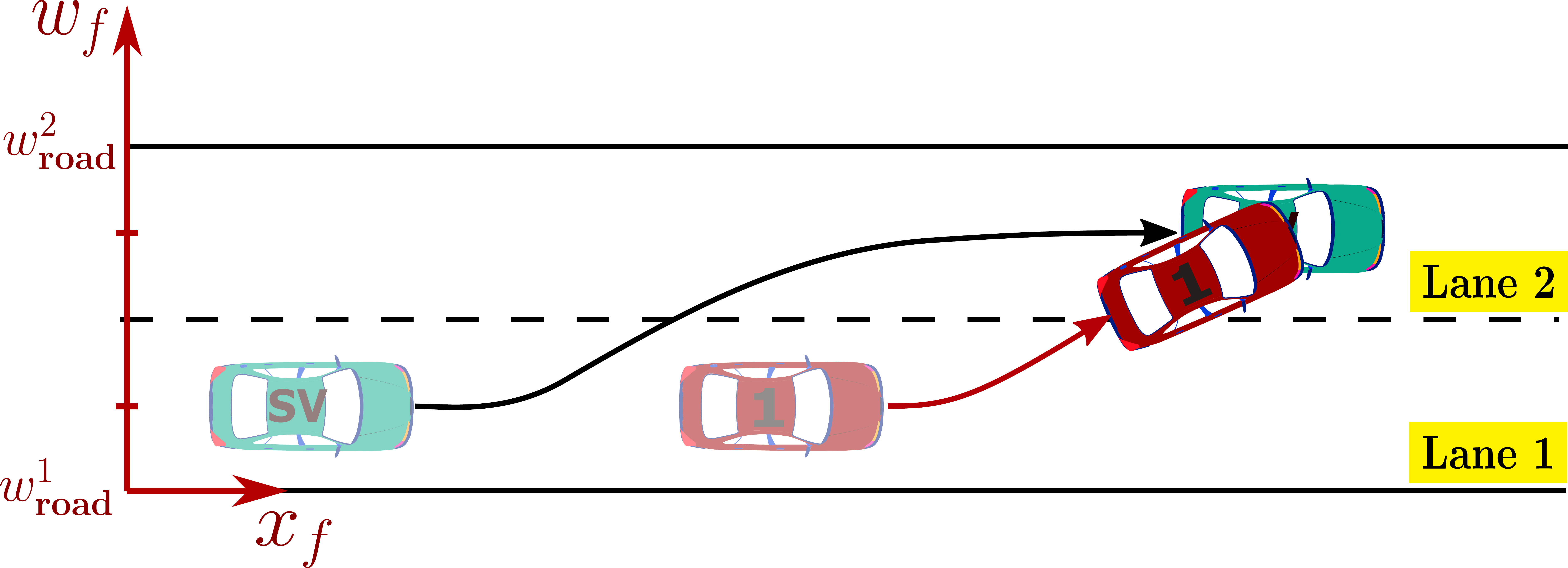}
  \caption{Collision cases for Logical~Scenario~2.}
  \label{fig:collision_log_sec_2}
\end{figure}

After 100 simulated experiments, GLIS identifies 9 collision cases. Three sample experiments with a collision are shown in Table~\ref{tab:num_test_results_logSec_2}. By analyzing the results, the collision triggering conditions are identified: ($i$) a combination of a relatively large $x_{f1}^0$ with a relatively small $v^0_1$ and a $t_c < t_{\text{exp}}$; ($ii$) a larger $x_{f1}^0$ is coupled with either a smaller $v^0_1$ or a lager $t_c$ or both. The specific values of $x^0_1, v^0_1$, and $t_c$ varies since they are correlated, while the collision triggering combinations all have the following features. Initially, SV changes lane to avoid OV$_1$ on Lane~1, then SV collides with OV$_1$ after $t_c$ on Lane~2 during lane-changing of OV$_1$ (Fig.~\ref{fig:collision_log_sec_2}). 
The collision is not avoidable since SV does not have enough response time to decelerate for the sudden lane-changing of OV$_1$. Additionally, lane-changing is not an option for SV to avoid OV$_1$. To eliminate these collision cases, one has to refine the ODD definitions (\emph{e.g.}, increase the lower bound of $v^0_1$, include another lane). Upgrading the MPC controller of SV without significant changes to the design is not feasible unless one also incorporates obstacle-avoidance mechanisms for the MPC controller of OV$_1$.

\begin{table}[!bt]
\caption{Numerical Test - Results (Logical~Scenario~2)}\label{tab:num_test_results_logSec_2}
\begin{adjustbox}{width=0.48\textwidth}
\begin{tabular}{l|ll}
\hline
Test               & Iter & $x_{\text{scene}} = [x^0_{f1},v^0_{1},t_c]$                                   \\ \hline

\multirow{2}{*}{1} & 28   & [12.57	46.94	16.75]'  \\
                   & 16$^*$   & [17.53	47.48	23.65]'\\
                   & 88   & [44.54	41.26	16.02]'\\ \hline

\multicolumn{3}{l}{Three sample iterations that can lead to collision are shown on the table.}\\
\multicolumn{3}{l}{(The ones with  $^*$ are the `best'/most critical ones identified by the optimizer }\\
\multicolumn{3}{l}{among these collision scenarios.)}\\
\end{tabular}
\end{adjustbox}
\end{table}

\section{Conclusion} \label{sec:conclusion}
In this paper, we investigated the application of the optimization method GLIS for computing safety-critical test scenarios of a designed feedback control system in an AD vehicle. The global optimization framework based on learning a surrogate model of the criticality
function introduced in this paper could effectively determine safety-critical test scenarios in the considered case studies. The information obtained from the corner cases found can then be used to refine the ODD definitions and/or upgrade the design of the system. Although we focused on analyzing control systems in this paper, the proposed framework is general and can be applied to other AD software components, and to many other V\&V tasks in the automotive domain or in other areas. 

One difficulty in the proposed framework is the design of the objective function, which is required to guide the search. It is often based on multiple criteria, and the formulation can be hard to determine beforehand.
Current research is devoted to ease the process of objective function definition.

%
%
%
%


\bibliographystyle{IEEEtran} 
\bibliography{Biblio_crtiSen}

\end{document}